\documentclass[aps,prl,twocolumn,groupedaddress]{revtex4}

\def\({\left(}
\def\){\right)}
\def\[{\left[}
\def\]{\right]}

\begin{document}


\title{Do Mixed States save Effective Field Theory from  BICEP?}


\author{Hael Collins}
\email{hcollins@andrew.cmu.edu}
\author{R.~Holman}
\email{rh4a@andrew.cmu.edu}
\author{Tereza Vardanyan}
\email{tvardany@andrew.cmu.edu}
\affiliation{Physics Department, Carnegie Mellon University, Pittsburgh PA 15213, USA}


\date{\today}

\begin{abstract}
The BICEP2 collaboration has for the first time  observed the B-mode polarization associated with inflationary gravitational waves.  Their result  has some discomfiting implications for the validity of an effective theory approach to single-field inflation since it would require an inflaton field excursion larger than the Planck scale. We argue that if the quantum state of the gravitons is a {\em mixed} state based on the Bunch-Davies vacuum, then the tensor to scalar ratio $r$ measured by BICEP is different than the quantity that enters the Lyth bound.  When this is taken into account, the tension between effective field theory and the BICEP result is alleviated.
\end{abstract}


\maketitle

The BICEP2 results\cite{Ade:2014xna} on the observation of primordial B-mode polarization of the CMB have ushered in a new era in cosmology. We now know that there is inflationary physics at an energy scale $V^{1\slash 4}\sim 10^{16}\ {\rm GeV}$. Of equal and perhaps greater importance, at least from an inflationary model building point of view, is the related fact that an observable B-mode signal necessarily implies that the inflaton field $\phi$ had to execute a very large excursion in field space, $\Delta \phi\gtrsim M_{\rm Pl}$. This result follows from the so-called Lyth bound\cite{Lyth:1996im} which we review below. 

If this is really the case, then the treatment of inflation in terms of the evolution of a scalar field zero mode rolling down a potential is in serious theoretical trouble. We expect that if nothing else, $M_{\rm Pl}$ will serve as the ultimate cutoff for any effective field theory, in particular that of the inflaton. However, if we are indeed forced to require that $\Delta \phi\gtrsim M_{\rm Pl}$ during inflation, it becomes unclear whether we can think of the Lagrangian for the inflaton $\phi$ as consisting of a derivative expansion of operators suppressed by the Planck scale. In particular, unless the Wilson coefficients of {\em all} the higher dimension terms in the potential were unnaturally small, we would have to keep an infinite number of terms in the potential, thus destroying any predictive power of the theory. Furthermore, there would be no a priori reason to suppose that radiative corrections would not destabilize this tuning. 

It is possible that once the full UV complete theory of inflation is unveiled, the resolution of this problem will become clear. In particular, there are possible scenarios, \cite{Kaloper:2008fb,McAllister:2008hb,Salvio:2014soa,Ashoorioon:2009wa,Choudhury:2013iaa,Hotchkiss:2011gz} that might be able to give the required B-mode signal while keeping field excursions to be smaller than $M_{\rm Pl}$. In this Letter, we would like to explore an alternative scenario, where it is the initial quantum state of the tensor fluctuations that comes in to save the day. There is a large literature on the initial quantum state of {\em scalar} metric perturbations in inflation (for a small sample see refs.\cite{Martin:2000xs,Danielsson:2002kx,Meerburg:2009ys,Kaloper:2002uj,Burgess:2002ub,Collins:2005nu,Ashoorioon:2010xg}), but less has been done on tensor perturbations\cite{Hui:2001ce,Lello:2013awa,Brahma:2013rua,Ashoorioon:2013eia} no doubt due to the lack of data until now. We will adapt the results in refs.\cite{Agarwal:2012mq,Collins:2013kqa} to argue that if the tensor modes are initially in a {\em mixed} quantum state, the tension between the BICEP results and the effective field theory treatment of inflaton evolution can be alleviated. 

First, let's review the issues associated with the Lyth bound. Recall (see, for example ref.\cite{Weinberg:2008zzc}) that the scalar and tensor power spectra can be written as:

\begin{equation}
\Delta_S^2(k) = \frac{1}{8\pi^2} \left . \frac{H^2}{\epsilon M_{\rm Pl}^2} \right|_{k= a H},\ \Delta_T^2(k) = \frac{2}{\pi^2} \left . \frac{H^2}{M_{\rm Pl}^2} \right|_{k= a H},
\end{equation}
where $H$ is the Hubble parameter during inflation and $\epsilon = -\dot{H}\slash H^2$ is the usual slow-roll parameter. If we take the ratio of these quantities we find
\begin{equation}
r \equiv \frac{\Delta_T^2(k)}{ \Delta_S^2(k)}=16 \epsilon.
\end{equation}
We stop for a moment to emphasize the following point: these expressions for the power spectra are {\em specific} to the choice of the Bunch-Davies (BD) state\cite{Bunch:1978yq} for both the scalar and tensor fluctuations. 

On the other hand, from the Friedmann equations it follows that 
\begin{equation}
\epsilon = \frac{1}{2 M_{\rm Pl}^2} \frac{\dot{\phi}^2}{H^2}
\end{equation}
so that if we use the definition of the number of e-folds as $dN = H dt$, we arrive at:
\begin{equation}
\frac{1}{M_{\rm Pl}} \frac{d\phi}{d N} = \sqrt{2\epsilon}.
\end{equation}
This relation is independent of any choice of vacuum state for the fluctuations, to the extent that we are neglecting the backreaction of the fluctuations on the zero mode equation of motion. If we now {\em assume} that the BD state has been chosen, we can relate $\epsilon$ to $r$ to find
\begin{equation}
\frac{1}{M_{\rm Pl}} \frac{d\phi}{d N}=\sqrt{\frac{r}{8}}\Rightarrow \frac{\Delta \phi}{M_{\rm Pl}}=\sqrt{\frac{r}{8}} \Delta N.
\end{equation}
Here $\Delta N$ is the number of e-folds corresponding to when the observed scales in the CMB, say those with $1\lesssim l \lesssim 100$ leave the inflationary horizon. When all the numbers are put in, the result is that 
\begin{equation}
\frac{\Delta \phi}{M_{\rm Pl}} \gtrsim {\cal O}(1) \sqrt{\frac{r}{0.01}}.
\end{equation}
For the value of $r$ found by BICEP, it follows that the inflaton field had to execute an excursion at least as large as $M_{\rm Pl}$. 

As emphasized above, this result requires the relations between the power spectra that follow from the use of the BD vacuum state for {\em both} the scalar and tensor fluctuations. Suppose we were to relax this; what would change in the above analysis? 

First of all, we need to decide how much to relax the BD assumption. The PLANCK data\cite{Ade:2013uln} is in great agreement with the choice of the BD vacuum for the scalar perturbations and for simplicity, we use this result to fix the scalar initial state to be the BD state. Next, we allow for the tensor initial state to be a mixed state based on the BD vacuum. What this means is that we can write this initial state as a density matrix in field space\cite{Agarwal:2012mq} (here $\Phi$ denotes either of the polarizations of the tensor mode):
\begin{equation}
\rho\(\Phi^{+}, \Phi^{-}; t_{0}\) = N \exp\(i {\cal S}_2\[\Phi^{+}, \Phi^{-}; t_{0}\]\),
\end{equation}
with 
\begin{eqnarray}
& & {\cal S}_{2} \[ \left\{ \Phi_{\vec{k}}^{+} \right\}, \left\{ \Phi_{\vec{k}}^{-} \right\}; t_{0} \]=\nonumber\\
& &  \frac{1}{2} \int \frac{{\rm d}^{3} k}{\(2\pi\)^{3}} \ \Big\{ \Phi_{\vec{k}}^{+}(t_{0}) A_{k} \Phi_{-\vec{k}}^{+}(t_{0}) - \Phi_{\vec{k}}^{-}(t_{0}) A_{k}^{*} \Phi_{-\vec{k}}^{-}(t_{0}) \nonumber \\
& &  + \ i \Phi_{\vec{k}}^{+}(t_{0}) B_{k} \Phi_{-\vec{k}}^{-}(t_{0}) + i\Phi_{\vec{k}}^{-}(t_{0}) B_{k} \Phi_{-\vec{k}}^{+}(t_{0}) \Big\}.
\end{eqnarray}
This is a Gaussian with the kernel $B_k$ indicating that there is mixing in this state. If we write the kernels as \cite{Berges:2004yj}
\begin{eqnarray}
-i A_{k} = \frac{\sigma_{k}^{2} + 1}{4 \xi_{k}^{2}} - i \frac{\eta_{k}}{\xi_{k}}, \quad B_{k} = \frac{1 - \sigma_{k}^{2}}{4\xi_{k}^{2}},
\label{eq:kernelscorrs}
\end{eqnarray}
then $\xi_{k}^{2}$ is the two-point function $\big\langle \Phi_{\vec{k}} \Phi_{-\vec{k}} \big\rangle (t_{0})$, $\xi_{k} \eta_{k}$ is the (symmetrized) correlator between the field $\Phi_{\vec{k}}$ and its conjugate momentum $\pi_{-\vec{k}}$ at the initial time, while the combination $\eta_{k}^{2} + \sigma_{k}^{2} / 4 \xi_{k}^{2}$ is the momentum-momentum correlator. The parameter $\sigma_{k}$ is a measure of how mixed the state is: ${\rm Tr}[\rho^{2}(t_{0})] = \Pi_{k} \( 1/\sigma_{k} \) \leq 1$, with equality only for a pure state. 

For our purposes, the most important aspect of mixed states is that when we compute the two point function including the quadratic terms coming from the initial state, we find (see the appendix in ref.(\cite{Agarwal:2012mq})) that the equal time correlation functions are {\em enhanced} by the factor $\sigma_k$:
\begin{equation}
\langle \Phi_{\vec{k}} \Phi_{-\vec{k}}\rangle(t)\equiv {\rm Tr}\(\rho(t) \Phi_{\vec{k}} \Phi_{-\vec{k}}\) =\sigma_k \left . \langle \Phi_{\vec{k}} \Phi_{-\vec{k}}\rangle(t)\right|_{\sigma_k=1}.
\end{equation}
Assuming the same $\sigma_k$ for both polarizations of the tensor fluctuations for simplicity, we now have

\begin{equation}
r =  \frac{\Delta_T^2(k)}{ \Delta_S^2(k)} = \sigma_k \left( \frac{\Delta_T^2(k)}{ \Delta_S^2(k)}\right)_{\rm BD\ vacuum} = 16\sigma_k \epsilon.
\end{equation}

Following this through, the Lyth bound now becomes
\begin{equation}
\frac{\Delta \phi}{M_{\rm Pl}}=\sqrt{\frac{r}{8 \sigma_k}} \Delta N \simeq {\cal O}(1) \sqrt{\frac{r}{0.01 \sigma_k}}.
\end{equation}
We see that the constraint of having a mixed state goes in the right direction to reduce the necessary excursion size for $\phi$, since $\sigma_k$ is necessarily greater than $1$. In particular, $\sigma_k\gtrsim {\cal O}(10)$ would suffice to keep $\Delta \phi <M_{\rm Pl}$.

Once we change the state from the BD one there is always the question of whether the backreaction of the energy density in the new state is small enough to allow inflation to proceed. This has been dealt with extensively for scalar perturbations, and exactly the same considerations apply here. In essence this requirement will restrict the (very) high-$k$ behavior of $\sigma_k$, but will not have a significant impact on $\sigma_k$ for the scales observed by either PLANCK or BICEP.

One potential issue with our result is what happens to the already existing tension between the bounds on $r$ from PLANCK\cite{Ade:2013uln} and the BICEP result when we allow for a mixed state to appear for the tensors. This requires more a detailed analysis, but the BICEP collaboration has already argued that this conflict may be avoided if running of the scalar spectral index is allowed. The mixing parameter $\sigma_k$ will also run in general, and it may be that the combination of these effects can render the results consistent with each other while still avoiding the problems coming from the Lyth bound. We are currently examining these issues.

The spectacular BICEP results have set up a clash between the description of inflation as being driven by a single field rolling down a flat potential and the restrictions coming from effective field theory. It is certainly early days yet in the field of B-mode cosmology, but assuming the BICEP results stand the test of time, this conflict must be resolved. It may well be that ideas from string theory, say, may take care of the problem. However, we have approached the problem from a different perspective. There is a simplicity to the choice of the BD state as the correct vacuum state of inflationary fluctuations, but there is no reason to suppose it is the {\em only} choice that could be made. Once an initial time, such as that of the onset of inflation is chosen, there are many possibilities consistent with both the data and field theoretical restrictions. What we have seen is that the choice of a mixed state based on the Bunch-Davies state can reduce the size of the field excursion necessary to achieve $r=0.2$ as seen by BICEP. While we are not necessarily advocating that the BICEP results be treated as the observation of a mixed state for tensors, we would argue that we need to approach this result with an open mind.

\begin{acknowledgments}
R.~H. and T.~V. were supported in part by the Department of Energy under grant DE-FG03-91-ER40682, as well as by a grant from the John Templeton Foundation. 
\end{acknowledgments}

\bibliography{bicep}

\end{document}